\documentclass{appolb}
\usepackage{amsmath}
\usepackage{graphicx}

\begin{document}
\title{The role of the kinematical constraint and non-linear effects in the CCFM equation%
\thanks{Presented at Excited QCD 2015, Tatransk\'a Lomnica, Slovakia}%
}
\author{Michal Deak
\address{IFIC, Universitat de Val\`encia-CSIC,\\Apt. Correus 22085, E-46071 Val\`encia, Spain}}

\maketitle
\begin{abstract}
We report on recent study~\cite{Deak:2015dpa} of the role of the kinematical constraint in the CCFM equation and its non-linear extension. We compare numerical results obtained by solving the CCFM equation and argue that kinematical constraint represents an important correction.\end{abstract}
\PACS{PACS numbers come here}

\section{Introduction}
In the regime of high energy scattering, where the center 
of mass energy is larger than any other available scale, perturbative approach 
to processes with high momentum transfer allows factorization 
of the cross section into a hard matrix element with initial off-shell gluons 
and an unintegrated gluon 
density~\cite{Gribov:1984tu,Catani:1990eg}. The unintegrated gluon density is a 
function of the 
longitudinal momentum fraction $x$ and transverse momentum ${k_T}$ of a gluon.
After taking into account formally subleading corrections coming from coherence 
of gluon emissions one is lead to the CCFM equation
which introduces gluon density dependent on hard scale related to probe. In principle, it is an equation that should be the ideal framework for application to final states at high energies and covering DGLAP and BFKL domain in gluon channel. It has been implemented in the Monte Carlo event generator \cite{Jung:2010si}.
However, so far good agreement with high precision data has been successfully achieved only in rather inclusive processes like  $F_2$ and Drell-Yan \cite{Hautmann:2013tba}. It is known that on the theory side the CCFM physics is still to be completed. For 
instance:
the impact of the kinematical effects introducing energy conservation in the CCFM evolution 
have been not investigated in all detail. As it turns out from our study it is necessary to 
revisit the inclusion of the so called kinematical constraint~\cite{Kwiecinski:1996td} into CCFM equation. 
Only recently the CCFM has been promoted to non-linear 
equation~\cite{Kutak:2011fu,Kutak:2012yr,Kutak:2012qk} therefore allowing 
for the possibility to investigate interplay of coherence effects and 
saturation \cite{Gribov:1984tu}.
In particular the important question is what is the role of the angular ordering 
and kinematical effect in the evolution at the non-linear level. The optimal 
form of initial conditions is also not known.   

We present here a continuation of the work done in~\cite{Deak:2012mx,Kutak:2013yga}.

\subsection{The CCFM equation and the kinematical constraint}
\label{sec:ccfm}
The CCFM equation for gluon density reads:
\begin{equation}\label{eq:IS-CCFM}
\begin{split}
\mathcal{A}(x,{k_T},p)=\mathcal{A}_0(x,{k_T},p)+{\bar\alpha}_S
\int&\frac{d^2\bar{\bf q}}{\pi\bar{\bf q}^{2}}
\int\limits_{x}^{1-\frac{Q_0}{\bar q}}dz\;
\theta(p-z{\bar{q}})\;{\mathcal P}(z,{k_T},\bar{q})\times\\
&\Delta_S(p,z\bar{q})\; \mathcal{A}(x/z,{k_T}^{\prime},\bar{q}),
\end{split}
\end{equation}
where $k^\prime=|{\bf k}+(1-z){\bf \bar q}|$ and the modulus of two dimensional vectors transversal to the collision plane are denoted $|{\bf k}|\equiv k_T$, $|{\bf q}|\equiv q_T$ and $x$ is gluon's longitudinal momentum fraction and $\bar\alpha_s=N_c\alpha_s/\pi$.
Also the rescaled momentum is introduced as $\bar{q}=|\bar{\bf q}|={q_T}/(1-z)$.\\
Where ${\mathcal P}\left(z,{k_T},\bar q\right)$ is the gluon splitting function.
The function $\Delta_S(p,z\bar{q})$ is the Sudakov form-factor.
The non-Sudakov form-factor $\Delta_{NS}(z,k_T,\bar{q})$ regularizes $1/z$ singularity.

\subsection*{The kinematical constraint}\label{sec:kc}
The integration over $\bar{\bf q}$ in the equation~\eqref{eq:IS-CCFM},
although being constrained from below by the soft cut-off $Q_0$, is not 
constrained by an upper limit thus violating the energy-momentum conservation. 
Moreover in the low $x$ formalism one requires that in the denominator of the 
off-shell gluon propagator one keeps terms that obey $|k^2|=k_T^2$.
In order to be consistent the non-Sudakov form-factor 
should be accompanied by a kinematical constrained limiting the above 
integration over $\bar q$. In approximated form it reads ${k_T^2}>z\,\bar q^2$
and at $z\ll 1$ guaranties that $|k^2|\simeq k_T^2$.
In \cite{Kwiecinski:1996td} it has been extended to region including also the case when $z\sim 1$. Careful derivation leads to following form of the kinematical constraint $k_T^2>\frac{z\,{q_T}^2}{1-z}=\,z\,(1-z)\,{\bar q}^2$.
The lower bound on $z>x$ results in the upper bound on 
$q_T^2<k_T^2/x\simeq{\hat s}$ providing local condition for energy-momentum 
conservation.
We include the kinematical constraint in the CCFM equation.
The non-Sudakov form-factor after inclusion of the full form of the 
kinematical constraint assumes form
\begin{equation}
\label{eq:nonsud2}
\begin{split}
\Delta_{NS} (z,k_T, \bar{q})= \exp \bigg\{ -
\overline{\alpha}_S & \int_z^1 \frac{dz^\prime}{z^\prime} \; \Theta
\left ( \frac{(1-z^\prime)k_T^2}{(1-z)^2\bar q^2} - z^\prime \right )\times\\ \: &\int \:
\frac{d{q}^{\prime 2}}{{q}^{\prime 2}} \; \Theta (k_T^2 -
{q}^{\prime 2}) \; \Theta ({q}^{\prime} - z^\prime\bar q) \bigg\}\;.
\end{split}
\end{equation}
Please note the presence of the function $\theta\left(\frac{k_T^2}{(1-z)\bar{q}^2}-z\right)$. 
The authors of \cite{Kwiecinski:1996td} solve the CCFM at small $z$ limit therefore the function $\theta\left(\frac{k_T^2}{(1-z)\bar{q}^2}-z\right)$ is neglected and in most of the  phenomenological and theoretical applications of the CCFM this term is neglected 
\cite{Hautmann:2013tba,Deak:2010gk,Salam:1998cp,Bottazzi:1998rs,Salam:1999ft,Avsar:2010ia,Bacchetta:2010hh,Chachamis:2011rw}.
The following form of non-Sudakov form factor is usually being used:
\begin{eqnarray}
\label{eq:nonsud11}
\Delta_{NS}(z,k_T,\bar{q}) & = & \exp \left ( - {\overline \alpha}_S
\int_z^{z_0} \frac{dz^\prime}{z^\prime} \int \frac{dq^{\prime 2}}{q^{\prime 2}}
\Theta
({k}^2 - q^{\prime 2}) \Theta (q^{\prime} - z^\prime \bar q) \right ) \label{a8}\\
& = & \exp \left ( - {\overline \alpha}_S \log \left (
\frac{z_0}{z} \right ) \log \left ( \frac{{k}^2}{z_0z{\bar q}^2} \right
) \right )\nonumber,
\end{eqnarray}
\noindent where
$z_0 = 1 $, if $({k_T}/{\bar q}) \geq 1$; $
z_0 = {k_T}/{\bar q}$, if $z < ({k_T}/{\bar q}) < 1$; $ z_0 = z $, if $({k_T}/{\bar q}) = z$. 
The discussed above $\theta$-function is not taken into account.

\subsection*{Saturation effects and kinematical constraint combined}
To account for gluon recombination at large gluon densities the CCFM equation 
has been promoted to non-linear equation by including a quadratic term \cite{Kutak:2011fu,Kutak:2012yr,Kutak:2012qk} and it 
reads:

\begin{align}\label{eq:IS-KGBJS}
\mathcal{A}&(x,{k_T},p)=\mathcal{A}_0(x,{k_T},p)+{\bar\alpha}_S
\int\frac{d^2\bar{\bf q}}{\pi\bar{\bf q}^{2}}
\int\limits_{x}^{1-\frac{Q_0}{\bar q}}dz\,\theta\left
(\frac{k_T^2}{(1-z)\bar{q}^2}-z\right)\theta(p-z{\bar{q}})\nonumber\\&{\mathcal P}(z,{k_T},{\bar q})\;\Delta_S(p,z\bar{q})\left(\mathcal{A}\left(\frac{x}{z},{k_T}^{\prime},\bar{q}\right)-
\delta\left(\bar{q}^2-{
\bar k_T}^2\right){\bar{q}}^2\;\mathcal{A}^2\left(\frac{x}{z},{\bar{q}},\bar{q}\right)\right),
\end{align}
where ${\bar k}_T=k_T/(1-z)$ and we included the kinematical constraint of the form \eqref{eq:nonsud2} 
in the kernel.
Simpler versions of the equation above have been already analyzed in~\cite{Kutak:2013yga} 
and it has been observed, that the equation leads to phenomenon called saturation at the saturation scale \cite{Kutak:2013yga,Avsar:2010ia}
and the saturation strongly suppresses the gluon density at low $x$ and low
$k_T$.
The natural question arises how are these results modified when some of the 
approximations are not taken and how are they modified  if the 
kinematical effect is imposed in the full form.

\section{Numerical results}

\begin{figure}[t!]
\vspace{0.5cm}
  \begin{picture}(30,0)
    \put(5, -105){
      \includegraphics{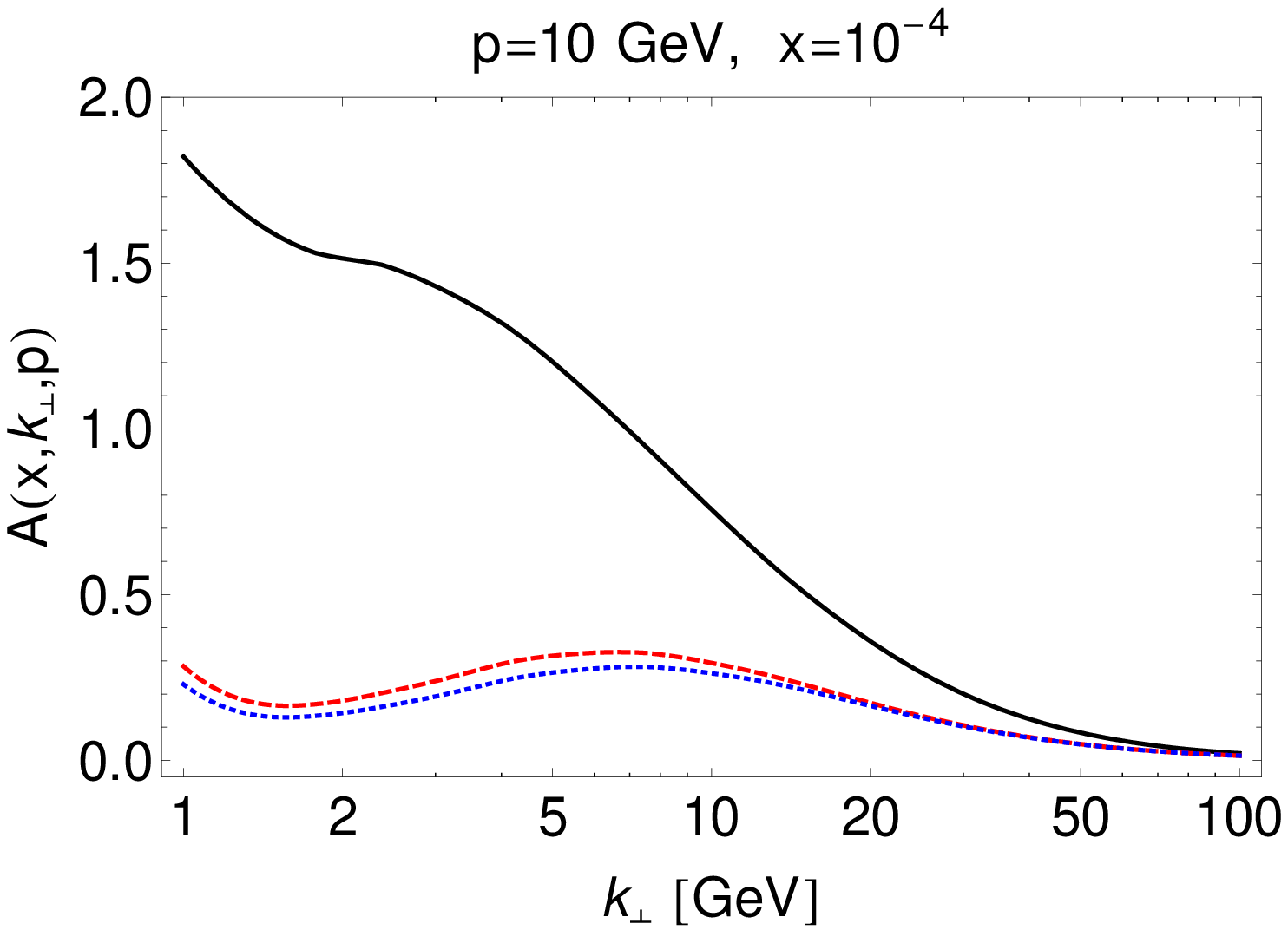}
    }    
    \put(189, -105){
      \includegraphics{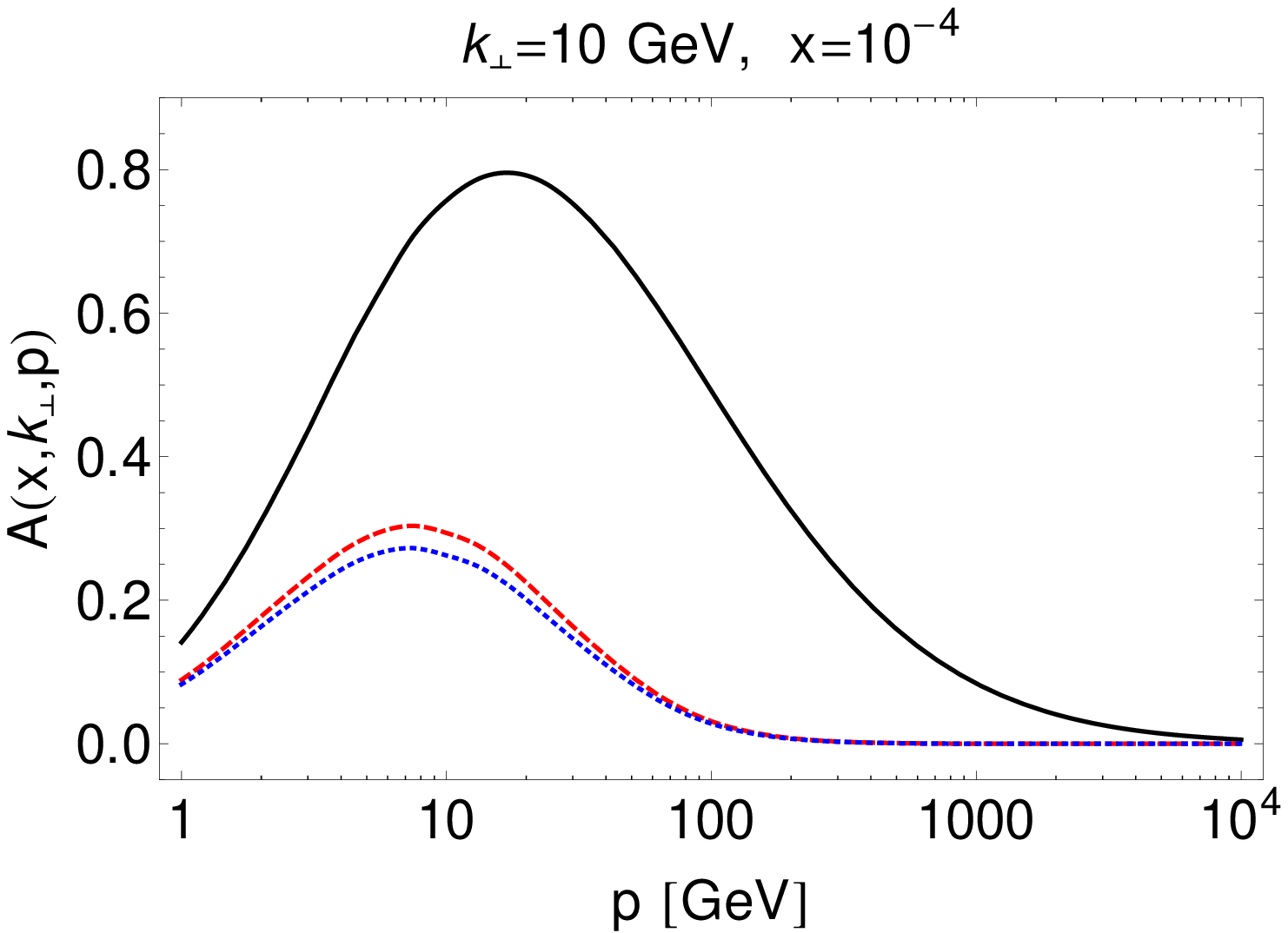}
    }
          
     \end{picture}
\vspace{3.6cm}
\caption{Relative ratio of CCFM and KGBJS 
solutions. Distributions with definite $p$ for varying value of $k_T$.}
\label{fig:plots1D-2}
\end{figure}

\begin{figure}[t!]
\vspace{1.5cm}
  \begin{picture}(30,0)
    \put(-8, -105){
      \includegraphics{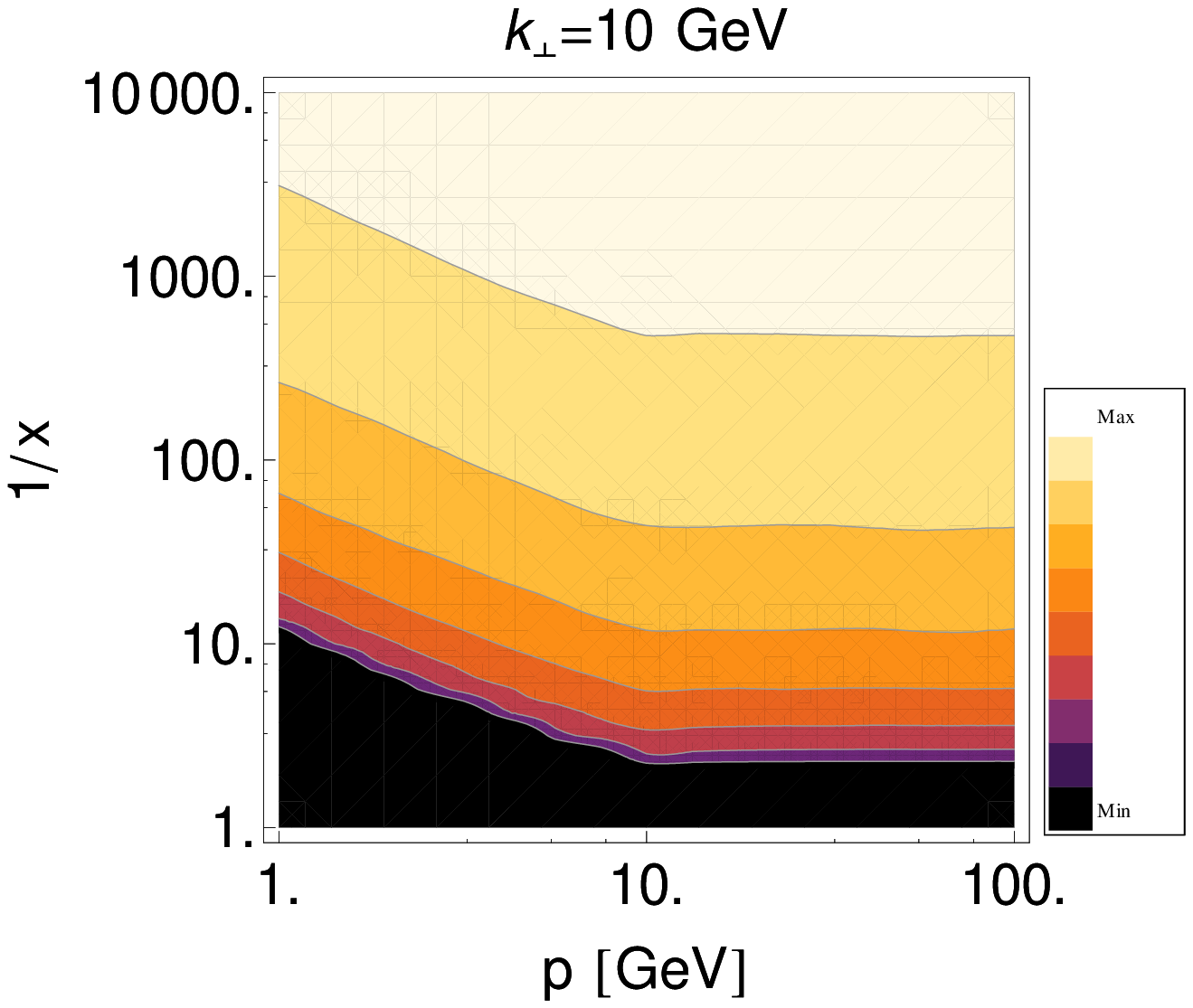}
    }    
    \put(174, -105){
      \includegraphics{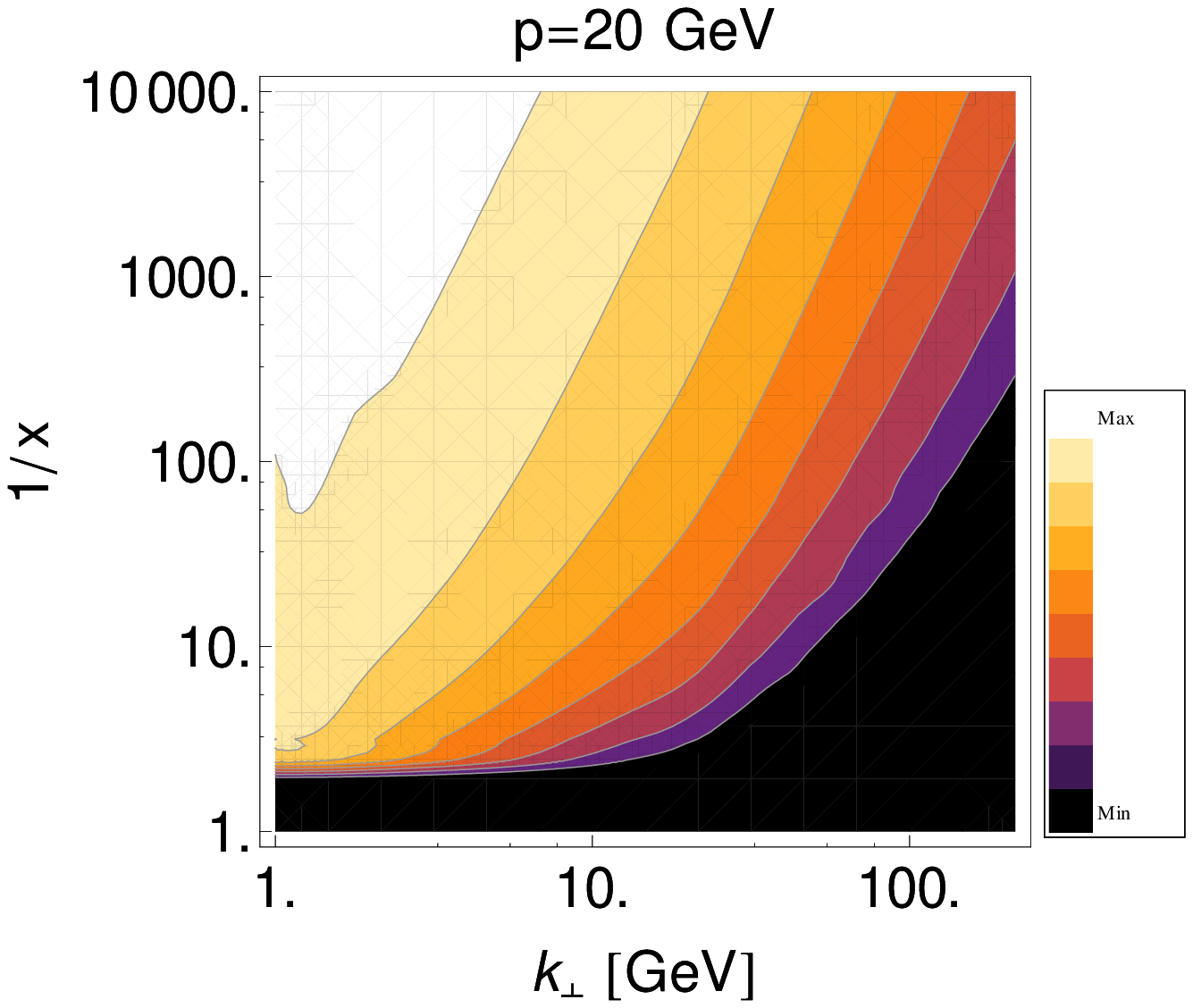}
    }
          
     \end{picture}
\vspace{3.6cm}
\caption{Relative ratio of CCFM and KGBJS 
solutions. Distributions with definite $p$ for varying value of $k_T$.}
\label{fig:plots2D-2}
\end{figure}

\subsection*{Linear equations}
We use 
an initial condition which includes resummed virtual and unresolved contributions, 
according to~\cite{Kutak:2013yga} and~\cite{Bacchetta:2010hh}, in the form 
${\mathcal
A}_0\left(x,k_T,p\right)=A\,\Delta_R(x,k_T)\,\Delta_S(p,Q_0)/k_T$,
with $A=1/2$ and 
$\Delta_R(z,k_T)$
is the Regge form-factor.

The observation we make from the plots (like plots in
Fig.~\ref{fig:plots1D-2}) is that the solutions of
equations we study differ significantly. The solutions exhibit also similar features. 
Solutions of both versions of the kernel with kinematical constraint exhibit a local 
maximum as functions of $k_T$ and $p$.
The positions of local maxima in the plots of $p$ dependence are correlated with the value
of $k_T$, with a shift to higher $k_T$ for the solution of the equation with kinematical constrain $\theta$-function included. 
The peak can be explained by the fact that 
the contribution of the integral on the right hand side of the equaiton peaks at 
around $k_T\sim p$. The peak therefore is a result of presence of
$\theta\left(p-z{\bar q}\right)$ -- angular ordering condition.
Similar peaks are present also in the plots of $k_T$ dependence and resemble Sudakov suppression of $k_T$ scales of the order of $p$~\cite{Kutak:2014wga}. 
However, in the 
case of without the $\theta$-function it seems, that the position of the peak does not depend on the value of $p$.
The peak observed in solution
of the equation without kinematical constrain $\theta$-function is 'hidden' under the result of the evolution. We can conclude that
the peak in the $k_T$ dependence is an interplay result of inclusion of the explicit kinematical contraint via
$\theta$-function factor and the Sudakov effect.

\subsection*{Non-linear equation}
We set the parameter characterizing the strength of the non-linear term $R$ the value  $R=\sqrt{1/\pi}\;$GeV in the 
equation~\eqref{eq:IS-KGBJS}.
By comparing the CCFM and KGBJS equations we see, that the kinematical constraint suppresses the 
growth of the gluon so much that the non-linear effects enter only at very low $x$.
Observations made in previous paragraphs are confirmed in 2-dimensional plots
(Fig.~\ref{fig:plots2D-2}), where we plot absolute 
relative difference of two amplitudes, solutions of the CCFM and KGBJS 
equations, defined by the quantity
\begin{equation}
\beta\left(x,{k_T},p\right)=\frac{|{\mathcal A}_{CCFM}\left(x,{k_T},p\right)-
{\mathcal A}_{KGBJS}\left(x,{k_T},p\right)|}
{{\mathcal A}_{CCFM}\left(x,{k_T},p\right)}\;.
\end{equation}
The function $\beta\left(x,{k_T},p\right)$, introduced before in~\cite{Kutak:2013yga}, can be used to measure the strength of the non-linear effects and to define a 
saturation scale using the conditions:
\begin{equation}\label{eq:conds}
\beta\left(x,Q_s\left(x,p\right),p\right)=const.{\rm ,} \quad \beta\left(x,k_T,P_s\right)=const.
\end{equation}
The second condition in~\eqref{eq:conds} defines $p$-related saturation scale.

The conditions above can be seen as equipotential lines in the plots in Fig.~\ref{fig:plots2D-2}, where different equipotential lines correspond to different constants on the right-hand side of the equation above.
The change in the slope of the $\beta\left(x,Q_s\left(x,p\right),p\right)$ at
around $k_T=p$ reported in~\cite{Kutak:2013yga}, can be seen 
clearly in Fig.~\ref{fig:plots2D-2}, can be
understood in the context of the peak at $p\sim k_T$ 
(Fig.~\ref{fig:plots1D-2}).

By comparing the plots Fig.~\ref{fig:plots2D-2} to analogous plots in \cite{Kutak:2013yga} we see that the main features are very similar. We therefore conclude that the $low-x$ approximation of the KGBJS and CCFM equations taken  in \cite{Kutak:2013yga} does not, at least, modify the relative difference between linear and non-linear equation.

\section*{Acknowledgements}
Michal Deak acknowledges support from Juan de la Cierva programme (JCI-2011-11382). This work has been supported in part by the Spanish
Government and ERDF funds from the EU Commission
[Grants No. FPA2011-23778, FPA2014-53631-C2-1-P
No. CSD2007-00042 (Consolider Project CPAN)].

\end{document}